\documentstyle[11pt,newpasp,twoside,epsf]{article} 
\def\arg#1{{\it#1\/}} 
\let\prog=\arg 
 
\def\edcomment#1{\iffalse\marginpar{\raggedright\sl#1\/}\else\relax\fi} 
\reversemarginpar 
\begin{document} 

\title{Quasar Atmospheres: Toward a `Low' Theory for Quasars}
\vspace{-0.5cm}
\author{Martin Elvis} 
\affil{Harvard-Smithsonian Center for Astrophysics, 60 Garden St.,
Cambridge MA 02138 USA}

\vspace{-0.3cm}
\begin{abstract} 
After 40 years we have no `low' theory of quasars to predict the
atomic absorption and emission features so abundant in quasar
spectra. Here I take the first step, using results selected for their
diagnostic power to build an `observational paradigm' unifying all the
atomic features in quasars. This paradigm bears a remarkable
resemblance to the one I proposed in Elvis (2000)! Yet all the results
used here were not used in that earlier study, with one exception.
The Elvis (2000) model has been tested by the new data and has
survived, and so has been strengthened. The structure is readily
tested, is physically suggestive, and opens up many areas for detailed
modelling.  From this I believe a truly predictive `low' theory of
quasars will emerge.
\end{abstract} 

\vspace{-1cm}
\section{INTRODUCTION: HIGH THEORY AND LOW} 
\vspace{-.3cm}

Enlightenment does not scale with published mass.  The quasar/AGN
literature has over 12,000 refereed published papers, and papers
now appear at a rate of about 2/day. Yet few would contend
that we have a much deeper understanding of quasars today than we did
30 years ago.  By 1974, just over 10 years after the discovery of
quasars, all of the current model for a quasar were in place: a
central massive black hole, an accretion disk and a relativistic jet.
We have made great observational progress mapping out quasar
properties at all wavelengths, and in the Unified Scheme (Padovani \&
Urry 1999) have resolved much of the confusing properties of quasars to
be due to the effects of obscuration.  But most quasar observations
are of atomic features.  The black hole/disk/jet theory is a `high'
theory, dealing only with overall energetics, and describes a naked
quasar, devoid of any of the veiling gas that creates these atomic
features. We need a `low' theory with sufficient detail to predict the
emission and absorption phenomenology of quasars. But to get to such a
theory in one step from 10$^4$ papers is not realistic. First we need
to build an `observational paradigm' to unify the phenomenology into a
coherent, over-constrained, and so robust, structure of what can
accurately be called the {\bf quasar atmosphere}.

\vspace{-0.2cm}
\section{THE BLIND MEN AND THE ELEPHANT: GETTING THE
QUASAR BIG PICTURE}

In the old Indian tale\footnotemark
\footnotetext{Best known in the West from the poem by J.D. Saxe. See
e.g.  http://www.noogenesis.com/pineapple/blind\_men\_elephant.html }
six blind men approach an elephant to understand what it might
be. The first happened on the elephant's flank and declared that an
elephant is a wall; the second found the tusk and pronounced an
elephant to be a spear; the third grasped the trunk, and decided an
elephant was a snake; the fourth felt a leg, and called an elephant a
tree; the fifth touched an ear and decided that an elephant is a
fan; and the sixth seized the tail and so said an elephant is a rope.
Each was confident in his diagnosis, {\em ``Though each was partly in
the right, And all were in the wrong!''}.  Moral: Not using all the
available information can be quite misleading. So when we study quasar
atmospheres we must not look at just one wavelength or feature, but
must consider all the features together (table~1).

\vspace{-0.5cm}
\begin{table}
\caption{Quasar Atmosphere Features}
\begin{tabular}{ll}
\tableline
Broad  Emission Lines  & BELs\\
\multicolumn{2}{l}{\em which we can divide into:}\\
~~High Ionization Broad  Emission Lines  & HIBELs\\
~~Low Ionization Broad  Emission Lines  & LOBELs\\
Broad  Absorption Lines& BALs\\
Narrow Absorption Lines& NALs\\
X-ray Warm Absorbers   & WAs\\
\multicolumn{2}{l}{Scattering phenomena (\em BAL troughs, VBELR,
Compton Humps, Fe-K)}\\
\tableline
\end{tabular}
\end{table}

\vspace{-0.7cm}
\section{A PARADIGM FOR QUASAR ATMOSPHERES}
\label{paradigm}

In Elvis (2000) I proposed a simple paradigm for quasar atmospheres.
To the high theory combination of black hole/disk/jet, I added a wind
arising from the accretion disk (figure~1).  The wind has specific
properties: coming from a restricted range of radii, initially making
a large angle to the disk plane (and so being roughly cylindrical),
and becoming radial well above the disk, making a $\sim$60\deg\ angle
to the disk axis to form a hollow bi-cone.

\begin{figure}[t]
\plotone{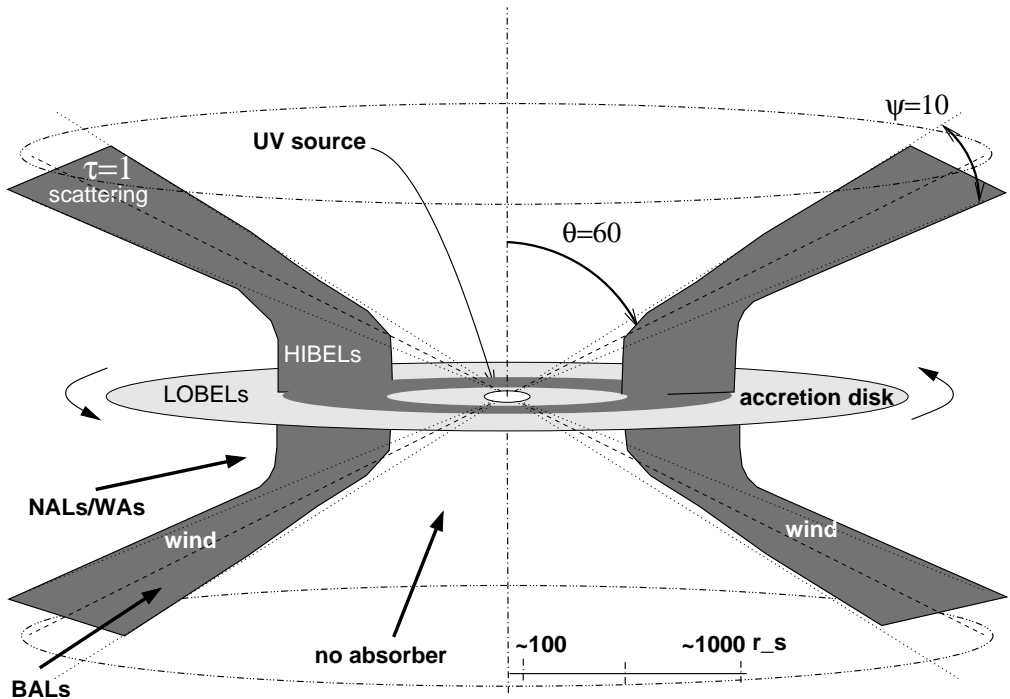}
\caption{\hspace{-0.5cm}A Paradigm for Quasar Atmospheres (Adapted from Elvis 2000)}
\end{figure}

Looking {\em through} the quasi-cylindrical part of the wind we
see NALs and WAs (Table~1). Because we are looking through a flattened
structure, scattering off the wind produces polarization.  When we
look near to the rotation axis our line of sight misses the wind
altogether and we see no absorption features. We are also looking down
on a highly symmetric structure, so polarization is weak.  A cool
phase of this wind emits the HIBELs. The LOBELs come from the outer
disk. As the wind becomes radial it is accelerated; looking down the
radial through this accelerating gas we see the BALs. The BAL wind is
marginally Thompson thick, producing the observed electron scattering
signatures.

This model is purely geometric and kinematic. At first this appeared
to be a weakness. We all want to go directly to the physics of this
wind. But let's recall that the Rees model for relativistic jets was
also just a geometric and kinematic model. To this day we do not have
an accepted theory of quasar jet acceleration, yet the simple model
allows us to understand quite a bit. Intermediate models or paradigms
such as these are valuable.  Note also that this model is independent
of the type~1/type~2 Unified Scheme. An obscuring torus at large radii
would simply shrink the opening angle derived for the bi-cone.

Several features of quasar atmospheres fall out as a natural
consequence of the structure proposed in Elvis (2000) [e.g. the BAL
covering factor; BAL `detachment velocities'; HIBEL blue-shifts.] This
is encouraging.  In the following I reconstruct the same paradigm
using only studies not used in Elvis (2000), except one.  The fact
that the structure can be re-built with almost none of the original
starting blocks is itself a strong test of the model and suggests that
there is something to it.

\vspace{-0.3cm}
\section{FILTERS: Finding Observations with Leverage}

Since there are so many papers on quasars\footnotemark\ we need a
means to filter out the many which do not point us toward a paradigm
\footnotetext{12,277 since 1963, from a ADS abstract search on 2003 April
18, for `quasar$\mid$AGN'.}
. I use three filters to accept papers: (1) {\em Physical
measurements}. e.g. Mass, length, density, (but {\em not} ratios,
column densities).
(2) {\em Absorption lines.} Since absorption only involves
an integration of the gas distribution over one spatial dimension, not
three, and because the sign of one of the three velocity dimensions is
also determined (i.e. blue-shift = outflow), which is not true for
emission lines.
(3) {\em Polarization.} Since any polarization implies a non-spherical
geometry and quasars, unlike stars, are not spherical even to first
order. (As we know from their jets.)


\vspace{-0.3cm}
\section{PHYSICAL MEASUREMENTS}

\vspace{-0.1cm}
\subsection{A Rotating, Large Scale-Height BEL Region}

Reverberation mapping of the BELs has been the most valuable technique
in quasar/AGN work. As a means of understanding quasar atmospheres the
most important result is the radius-line width relation for BELs
(Peterson \& Wandel 2000).  This correlation follows a Keplerian $v^2
r = constant$ relation, which tells us that the BEL gas is close to
bound and close to following simple orbits.  Peterson (2003, these
proceedings) demonstrates that there are still factors of 2-3
uncertainty in the normalization of $v^2 r$ to black hole mass, so
that pure Keplerian motion is not required.  However, the lack of
red-blue asymmetry in the reverberation response in NGC~5548 (Wanders
et al. 1995) already requires a large orbital component, as does the
polarization of H$\alpha$ (Robinson et al. 2002).

Variability of another kind - VLBI expansion - lets us derive another
physical measurement, the angle of the jet axis to our line of sight,
$\theta$, (figure~1, assuming we know $H_0$). Rokaki et al. (2003) use
this approach to derive jet angles for a sample of radio-loud quasars
for which they also measure H$\alpha$ FWHM and equivalent widths
(EW). Strikingly they find angles up to 40\deg, well away from the
region ($\theta<10\deg$) where a beamed continuum dominates the quasar
spectrum.  Rokaki et al. find that more edge-on objects have on
average broader FWHM(H$\alpha$), implying ordered rotation around the
jet axis. Similar correlations with radio core-to-lobe ratio, which is
a proxy for $\theta$, find the same result, but proxy measurements
cannot compare model predictions. Having $\theta$, Rokaki et al. can
do so, and they find that pure rotation is not a good fit, and that
another component of amplitude $\sim$2000~km~s$^{-1}$ is needed. We
return to this later (\S 6.3).

The EW(H$\alpha$) vs $\theta$ plot also shows a correlation. If
H$\alpha$ were from a flattened disk, as we expect the continuum to
be, then as we view the disk at larger and larger angles both would
suffer the same geometric and limb darkening, and so no change in EW.
Instead larger $\theta$ objects have larger EW(H$\alpha$).  Rokaki et
al. find that the EW vs. $\theta$ relation is just as expected if the
continuum comes from a disk and suffers cos~$\theta$ and limb
darkening, but the H$\alpha$ is isotropic.  So the BEL is not only
rotating, it also has a large scale height, rising well above the
continuum producing disk. I.e. a rotating cylinder.

A rotating cylinder is not an equilibrium configuration. A Keplerian
velocity on the disk becomes super-Keplerian above the disk, so gas
there will feel a radial centrifugal acceleration. A natural way of
producing this configuration is for BEL gas to be a wind rising off
the disk at a large angle to the disk.

\vspace{-0.2cm}
\subsection{A Limited Range of BEL Wind Radii}
\label{limited}

The reverberation mapping FWHM vs lag time correlation shows that the
BEL region is stratified, with HIBELs (Table~1), which have
FWHM$\le$10,000~km~s$^{-1}$, lying in the inner region at
$\sim$1000~$r_s$ ($r_s$=Schwartzchild radii). For NGC~5548 (Korista et
al. 1995) the ratio of the HeII to CIV lag times is roughly 3, so the
HIBEL region has an approximate thickness $\Delta r/r\sim$3, although
at this level it is not obvious how to define the correct lags
(Peterson 2003, these proceedings).

Collin-Souffrin et al. (1988) proposed that the LOBELs (Table~1) come
from a disk, while the HIBELs come from a large scale height flow.
Since the LOBELs are narrower and have longer reverberation times,
they lie outside the HIBEL region.  Once we think of the HIBELs as
coming from a wind, the Collin-Souffrin et al. picture requires that
the HIBEL wind arises only from a restricted range of radii.  This
idea is strongly supported by observations of two Narrow Line
Seyfert~1s (NLSy1s) by Leighly \& Moore (2003). They find that the
LOBEL MgII is narrow and symmetric, as implied by the NLSy1 label, but
that the HIBEL CIV is broad and asymmetric to the blue. This is
readily explained if the CIV velocity comes from a wind, with the far
(red-shifted) side of the wind obscured by the disk.  Similar
blueshifts of HIBELs relative to LOBELs have been known for a long
time in ordinary quasars (Gaskell 1982) but have remained
puzzling. These can now be understood as less extreme examples of the
effect found by Leighly \& Moore. In normal quasars the HIBELs are
dominated by disk rotation, but include blue-shifts of
$\sim$1000-3000~km~s$^{-1}$ due to a wind component. The lack of HIBEL
redshifts seems to put the wind at accretion disk dimensions, not on a
torus scale.

\vspace{-0.3cm}
\section{ABSORPTION}

\subsection{The NAL/WA gas: a 2-phase medium in Pressure Balance}
\label{2phase}

The idea that the NALs and WAs came from the same medium was
contentious only recently (Crenshaw 1997). Now that we have high
resolution (R$>$400) X-ray spectra of quasars from {\em Chandra} and
XMM-Newton, this concern has been laid to rest. The NALs and WAs both
have (1) narrow line widths, (2) similar outflow velocities of
$\sim$1000~km~s$^{-1}$ (Collinge et al. 2000), (3) closely related
ionization states (Krongold et al. 2003), and (4) occur in the same
objects so reliably that one can predict the other (Mathur et
al. 1998). So now we can use this equivalence to learn more about the
UV/X-ray revealed winds from quasars.

Krongold et al. (2003) find 2 ionization states in the X-ray
absorption spectrum of NGC~3783, and can exclude an intermediate third
state. This points away from a many-phase medium (Krolik \& Kriss
2001). The Krongold et al. model requires only 6 free parameters to
fit over 100 X-ray absorption features (75\% of which are
blends). Occam's razor says there is something to their
model.  They find that the pressure in the two gases is equal,
assuming only that they lie at the same distance from the central
continuum, as is suggested by their shared kinematics.  When the two
phases are plotted in temperature, T~vs~$\Xi$ (the ratio of radiation
pressure to gas pressure) space they lie on the equilibrium curve for
a gas illuminated by a continuum similar to that of NGC~3783.  Netzer
et al. (2003) find a 3rd, higher ionization, absorber and this may lie
on the upper branch of the equilibrium curve.  Krongold et al. (2004)
have now found a similar solution for another object, NGC~985, so the
result may be general.  The cool phase has T$\sim$10$^4$K, comparable
to the BEL gas, and also has a similar ionization parameter,
$log~U\sim$ -0.8. Hence a simple 2-phase medium in pressure balance is
strongly indicated. Only an extraordinary coincidence could be invoked
to explain the equal pressures if separate media are involved. The
geometrical relation of the 2 phases is unknown: e.g. is it layered or
foam-like.

\subsection{The NAL/WA gas: a Transverse Wind, Close to the Continuum}
\label{transverse}

How is the NAL/WA wind flowing? There is strong evidence that the
primary motion is transverse to our line of sight; so the true speed
of the wind is larger than the observed values. Arav et al. (2002)
note that the CIV doublet ratio in NGC~5548 is not 2:1 as required by
atomic physics. The only way out is to have extra continuum fill in
the base of the absorption lines. But Arav et al. also find that the
CIV doublet ratio varies systematically with velocity across the
lines, so scattering into the line of sight is untenable. The only
remaining possibility is that the absorbing gas covers different
fractions of the continuum source at different velocities. A radial
wind cannot do that unless our line of sight is very special. A
transverse wind can naturally incorporate velocity-dependent covering
factors if the wind accelerates across the diameter of the source.
This requires the wind to be quite close to the continuum source, say
within $\sim$10 continuum diameters (so that the continuum source
subtends $>$10\deg) since at larger distances: (1) partial covering is
improbable, (2) significant acceleration in a random small section of
the wind is unlikely. A systematic change in covering factor with
velocity also points to sheets rather than to a mist of clouds.

So the NAL/WA gas must lie in a transverse wind, close to the
continuum, at $\le$1000$r_s$ if the UV continuum comes from
$\sim$100$r_s$. [$T=3\times 10^4$K for $L=0.05~L_{Edd},
M_{\bullet}=10^8~M_{\odot}$ (Frank, King \& Raine 1992, {\em Eq.7.18})
for NGC~5548 parameters (Peterson \& Wandel 2000).]  Limited WA
variability monitoring gives similar sizes from recombination times
(Nicastro et al. 1999).  A transverse flow also implies a plane of
origin and a conical or cylindrical structure.

\vspace{-0.2cm}
\subsection{One Wind: Combining HIBELs and NALs/WAs}
\label{onewind}

In the above sections we have seen that the BEL is a wind, as is the
NAL/WA. We also saw that the BEL and the NAL/WA winds both: (1) rise
from a disk at a large angle; (2) lie roughly at $r \sim 1000~r_s$;
(3) have the same pressure; (4) have complementary ionization
parameters; (5) have non-orbital velocities of
$\sim$1000-3000~km~s$^{-1}$.  In addition the BEL wind is rotating,
something we do not know for the NAL/WA wind. The obvious simplest
explanation is that the BEL is just the cool phase of the NAL/WA
gas. There is only one disk wind from quasars, and it accounts for all
the atomic features discussed so far.

\vspace{-0.5cm}
\section{POLARIZATION: A THOMPSON THICK SCATTERER}

\subsection{The NAL/WA Wind is Flattened and Universal}
\label{nalwapol}

The optical polarization of AGN with NALs is higher than that of
non-NAL AGN (Leighly et al. 1997).  In a plot of \% polarization
vs. $log~N_H$ (from ASCA X-ray spectra) Leighly et al. find a clear
separation between low $N_H$ non-NAL objects (with low polarization,
$<$1\%), and high $N_H$ NAL objects (with high polarization,
$\sim$1-5\%). The scatterer in the NAL objects must be highly
flattened and seen close to edge-on. If follows that there must be
objects from the same population seen pole-on, but these objects
cannot be obscured by a large $N_H$ as there are no high $N_H$/low
polarization objects. So the scatterer and the obscurer share a
flattened co-axial geometry. The simplest explanation is that the
scattering structure is a part of the NAL/WA (and by implication
HIBEL) wind, and all AGN have NAL/WA winds. Yet the NAL/WA wind $N_H$
is only $\sim$10$^{22}$cm$^{-2}$, $\sim$1\% of ${\large \tau}_T$, insufficient
to produce strong polarization.

\subsection{All Quasars have Flattened BAL Winds}

BALs have widths $\sim$3-10\%$c$, some 10 times those of NALs.  BALs
are only seen in $\sim$15\% of quasars (Tolea, Krolik \& Tsvetanov
2002). However BALs are of much wider importance in quasars because
the strong polarization of the BAL troughs (Ogle et al. 1998) implies
that BAL winds are highly flattened (Ogle 1998) and so must be present
in all, or most, quasars. (This is the argument also used in Elvis
2000.)

If BAL winds are present in all quasars, but in most cases do not
enter our line of sight, then non-BAL quasars should also show
evidence of scattering off high velocity material. Ogle (1999, PhD
thesis and unpublished preprint) uses a formalism for polarization
from scattering off axisymmetric structures, first developed for Be
stars, to show that a wide angle bi-cone (half opening angle $\sim
60$\deg), viewed from a random set of directions, can reproduce the
distribution of optical polarization seen in non-BAL quasar samples.
The implied universality of BALs in quasars and their 15\% occurrence
rate require the BAL bi-cone divergence, $\psi$, angle to be small.

BAL winds are then an integral part of quasar structure, and any model
of quasar structure must include them. But we just showed that NAL/WA
winds are universal in AGN. Is there a connection?

\vspace{-0.4cm}
\section{A BEND IN THE WIND: WA/NAL/HIBEL and BAL Unity}

If the NAL/WA/HIBEL wind is transverse and forms a bi-conical
structure (see \S\S 6.2, 7.1), then there must be some cases where we
look through the wind right down the flow direction. How many cases
there are will depend on the wind divergence angle, $\psi$.
$\psi$=10\deg (for $\theta$=60\deg, figure~1) gives a 15\% covering
factor for the central continuum source, so if this is the correct
divergence angle then BALs could be simply the WA/NAL/BEL wind seen
end on. The acceleration of the wind would show up in the broad
profiles of the BELs.  There is a problem with this picture. In a
simple bi-cone with a 60\deg opening angle the typical NAL velocities
will be $\sim$ 90\% of the BAL velocity, instead of the 10\% observed.

Why should BALs not simply be the high terminal velocity/high
luminosity tail of the NAL population? After all, BALs are typically
found in luminous quasars (at z$>$2 in order to see CIV BALs in
ground-based optical spectra), while no low luminosity (low z) objects
have NALs with widths $>$4000~km~s$^{-1}$ (Kriss 2001).  This is not a
conclusive argument, as low luminosity sample sizes are typically only
of order 20, so that only $\sim$3 BALs are expected. The larger
samples of FUSE AGN, and Sloan Digital Sky Survey (SDSS) faint quasars
at z$>$2 will test this idea more strongly. Meanwhile it is useful to
consider alternatives.

A curious feature of BALs is their `detachment velocity'.  The onset
of a BAL is usually sharp, but does not occur at the center of the
corresponding BEL, but offset to the blue by 0-4000~km~s$^{-1}$.
There are three potential explanations: (1) the BAL wind accelerates
impulsively leaving negligible $N_H$ from 0-4000~km~s$^{-1}$; (2) the
BAL flow is initially over-ionized as it begins to accelerate, so line
driving is weak (see \S 5.2) and the wind must be hydromagnetic. (3)
The BAL flow bends into our line of sight only when it has already
accelerated to some significant fraction of the BEL FWHM velocity.

The latter, geometrical, explanation is simple, and implies that low
velocity NALs must be observed, since a bend in the wind allows many
viewing directions to cut through a point in the wind at a large angle
to the wind flow direction (figure~1), picking out a single
velocity. If we accept a bend in the wind then we combine the
NAL/WA/HIBEL wind with the BAL wind by a simple geometry.

If this unification is real we would expect evidence for high ${\large
\tau}_T$ and BAL velocity material in low luminosity quasars
(=AGN). The non-variable `narrow' Fe-K X-ray emission lines, and their
corresponding Compton humps, show that ${\large \tau}_T\sim$1 material
is present in AGN far from the X-ray source (e.g. Chiang et al. 2000).
Polarization and variability studies of BELs reveal a ``Very Broad
Emission Line Region'' or VBELR. These broad faint wings are mainly
seen in H$\alpha$ and are polarized (Young et al. 1999).  This
indicates that they are scattered radiation off electrons moving at
about twice the FWHM(H$\alpha$).  BALs typically have twice the BEL
FWHM (Lee \& Turnshek 1995), suggesting that the scattering material
could well be the BAL gas.  Since the VBELs do not vary, a scatterer
larger than the HIBEL region, consistent with the conical BAL part of
the wind, is a simple explanation. SDSS spectra (Hall et al. 2002)
show that the occasional detachment velocities that lie redward of
the peak of the emission lines can be explained by a rotating BAL wind
close to an extended continuum source. Why the wind bends outward will
be discussed in \S 9.2.


\vspace{-0.3cm}
\subsection{Return to a Paradigm for Quasar Atmospheres}

With the above arguments we have constructed the same basic picture of
quasar atmospheres proposed by Elvis (2000, see \S 3), out of almost
entirely new results.  All the components of quasar atmospheres listed
in table~1 are included.  The model explains several features not
built in, which were previously just puzzling observed facts. The
Elvis (2000) model has survived numerous tests: any of the data used
above could have been inconsistent with the model, but did not turn
out to be. This paradigm makes predictions, allowing stronger tests.

\vspace{-0.3cm}
\section{THE BEGINNINGS OF A LOW THEORY OF QUASARS}
\label{low}

Given the growing solidity of the Elvis (2000) observational paradigm,
we can begin to explore the physical implications. The realization
that quasar atmospheres are not static but are winds, marks a big
change in viewpoint.  Winds not only point to specific physics for
quasar atmospheres, they also allow us to think about the constraints
they put on the base of the wind, i.e. the accretion disk. Winds also
force us to ask what happens to the wind material as it leaves the
immediate quasar environment. Exploration of these topics is only just
beginning and we sketch some developments below.

\vspace{-0.3cm}
\subsection{A 2-phase Photoionized Wind}
\label{2phase}

Winds allow the application of an old, but discarded, result (Krolik,
McKee \& Tarter (1981).  Krongold et al.'s (2003) findings that: (1)
the WAs lie on the (T, $\Xi$) equilibrium curve, (2) they are on
stable branches of that curve, and (3) they have the same pressure,
points clearly to a few (2-3) phase gas photo-ionized by the central
continuum source. But a static confining medium for cool BEL clouds
destroys the BEL clouds in less than a dynamical crossing time via
shear forces. The realization that if the hot and cool phases are
co-moving in a (non-spherical) wind then differential forces on the
two phases become a 2nd order effect, and that any particular gas
condensation is gone from the system in a dynamical time (Elvis 2000)
removes this problem. We are looking, as it were, through a candle
flame. A flame appears almost steady, yet is an ever changing flow of
gas.

\vspace{-0.3cm}
\subsection{Line Driven Winds, with Caveats}
\label{line}

Photoionization is clearly important for quasar winds. Is radiation
pressure then the driver for the wind? In particular, given the
ionization state of the wind, is line driving via UV and X-ray
transitions accelerating the wind, as in O-stars.  There is evidence
that this is so. Arav (1996) note that a signature of `line-locking'
between Ly$\alpha$ and NV is seen (the `ghost of Ly$\alpha$').
Murray and collaborators (Murray \& Chiang 1995, Murray et al. 1995,
Chiang \& Murray 1996) have developed the theory of line driven quasar
winds to explain BALs and BELs. Given the presence of some shielding
gas at small radii (`hitchhiking' gas), these winds arise naturally in
quasar conditions. Proga (2003) has made detailed hydrodynamical
calculations for radiation driven winds and shows clearly that disk
winds can be made, and can match BEL profiles.

There are some characteristics of the paradigm that are not quite in
agreement with these models: (1) The observations seem to require the
disk wind to be close to vertical for some large height above the
disk, in order to produce NALs/WAs, while the radiation driven wind
models tend to produce more equatorial winds. (2) Pure radiation
driven winds must reach the escape velocity before a gas element makes
more than half an orbit, else the gas remains on a bound orbit and
will re-enter. [The gas is highly supersonic, since
$v_{sound}\sim$100~km~s$^{-1}$, so no gas pressure support is
possible.] Yet the observations point to most BEL gas being dominated
by orbital velocities. (3) The observed wind extends over a factor of
only a few (2-3) in disk radius, while the models tend to have winds
arising over the whole disk surface.

Hydro-magnetic winds, which are common in protostellar disks, may
address the first two points (e.g. Blandford \& Payne 1982; Konigl \&
Kartje 1994), since the field can confine the gas longer and may allow
it to rise at large angles to the disk, and be orbital velocity
dominated. How does the magnetic pressure affect the pressure balance
of the hot and cool phases, and does this affect the (T, $\Xi$)
curves?

\vspace{-0.4cm}
\subsection{Why is the Wind Thin?}
\label{thin}

The observed wind extends over a factor of only 2-3 in disk radius,
yet neither sets of papers referred to above point to this as a
consequence of their models. In fact, though, a limited range of radii
may be natural in line driven winds.

Both Leighly \& Moore (2003) and Risaliti \& Elvis (2003) model line
driven winds and find that the change in ionization parameter, $U$,
and the available driving UV/soft-X-ray radiation, control where a
wind can arise. At small radii the gas is over-ionized and few atomic
transitions are available, so electron scattering dominates, but is
too weak to create a wind. However this `failed wind' does screen gas
in the next outer radii, lowering $U$ (and providing an explanation
for the hitchhiking gas). As $U$ decreases below a critical value, the
number of transitions rises very fast and so the opacity does too,
making radiation line and driving suddenly become $\sim$1000 times
stronger than electron scattering, driving a wind. This creates an
inner boundary to the wind. The wind absorbs most of the UV and
soft-X-ray continuum, so that by some larger radius, where ${\large
\tau}_{\tiny UV}\sim$1, the radiation available to drive the winds
falls quickly, creating an outer boundary to the wind. Low ionization
BEL gas, excited by harder X-rays, is found further out where, 
the bound LOBEL gas tracks the disk kinematics.

Risaliti \& Elvis (2003) explore the range of ($M_{\bullet},
\dot{m}/\dot{m}_{Edd}$) for which winds are possible. At
$\dot{m}_{Edd}$ all quasars, inevitably have winds, but below
$\dot{m}/\dot{m}_{Edd}\sim$0.1 it becomes hard to initiate a
wind. Super-Eddington winds [L$>$L(Edd)] are different from those
discussed in this paper, and are probably important during the black
hole building phase of quasar, where they may create the black hole
mass, bulge mass ($M_{\bullet}-\sigma_V$) relation (King 2003).

\vspace{-0.2cm}
\subsection{Working Back: Constraining Quasar Accretion Disk Physics} 

What imparts the initial vertical push to the wind gas at the
accretion disk? Proga (2003) and Risaliti \& Elvis (2003) use the
local radiation pressure in a Shakura \& Sunyaev (1972) disk. Since
quasars {\em do} have winds we can explore how far a disk can stray
from the $L_r \propto r^{15/8}$ Shakura/Sunyaev dependence and still
make winds. We could also search for other mechanisms, such as the
magnetic field recombination which drives the solar wind.

The spectrum from the hot inner edges of accretion disks are highly
model dependent and there are numerous candidate models (e.g. Janiuk
et al. 2003), including emission from the `plunging region' within the
innermost stable orbit (Krolik \& Hawley 2002). This part of the disk
will emit in the EUV, which is inherently unobservable.  We may be
able to determine the EUV SED shape by requiring the (T, $\Xi$)
equilibrium curve to go through WA points, and varying the EUV SED
until consistency is achieved.  We may then constrain accretion
physics close to the black hole.

\vspace{-0.2cm}
\subsection{Thinking Ahead: the Dusty Fate of the BEL gas}

If the BEL gas is constantly being ejected from quasars what happens
to it?  Elvis, Marengo \& Karovska (2003) find that the cool BEL gas
expands adiabatically in quasar winds and will inevitably form large
amounts of dust as it passes through the (P,T) region where dust is
formed in AGB stars.  Conditions in the gas are more benign than in
AGB stars.  Quasar winds could then be a major source of dust at high
redshifts.  The most luminous quasars with 10$^{47}$erg~s$^{-1}$ have
the highest dust masses, and may lose $>$10~M$_{\odot}$~yr$^{-1}$. For
the same 1\% dust fraction as in AGB stars this gives
$\sim$10$^{7}$~M$_{\odot}$ of dust over a 10$^{8}$~yr lifetime. This
approaches the detected amounts, but still short by a factor
$\sim$10. Super-solar abundances in high redshift quasars (Hamann \&
Ferland 1999) will provide a higher density of atoms to form precursor
molecules. Dust condensation is highly nonlinear so the amount of dust
formed is hard to predict, but initially the dependence on abundance
is likely to go as $n^2$, allowing more dust to be produced.  So the
infrared emission of quasars may not require the normally assumed
associated starburst.

\vspace{-0.3cm}
\section{CONCLUSIONS}
\vspace{-0.2cm}

I have presented a structure for quasar atmospheres. This structure is
simple, and increasingly well supported by the data. The structure
explains a variety of, once unrelated, quasar observations and puts
them into a self-consistent physical picture.  This picture is
physically suggestive, and gives us the prospect not only of
understanding quasar winds and atmospheres, but of working back to the
physics of accretion, and out to the impact of quasar winds on their
larger scale environment.
%
%
If the view presented here continues to fit the data, then we have the
beginnings of a low theory of quasars, starting with their
atmospheres. Quasar studies can then enter a stage in which they
connect with detailed physics and so link quasars to the rest of
astrophysics.

\vspace{-0.5cm}

\newpage
\appendix\section{POSTSCRIPT: IMAGING QUASARS}

All of the indirect reasoning given above to derive what a quasar
looks like is needed because quasars have always been point
(`quasi-stellar') sources. But we now know the length scale of quasar
atmospheres from reverberation studies of BELs.  Combining these with
the WMAP metric gives us their angular diameters. These turn out to be
$\sim$0.1~mas for nearby AGN and only 10 times smaller
($\sim$10~$\mu$as) for luminous high z quasars (Elvis, \& Karovska
2002, ApJL, 581, L67).

So an optical or near-IR interferometer of baseline $\sim$100~km could
resolve BEL structure in quasars and render this indirect argument
moot. Similarly a UV interferometer in space with a $\sim$10~km
baseline could image the nearby ($z<$2) HIBELs. Inverting this
approach, imaging reverberation mapping of high z quasars could map
out the metric to $z>$6, and independently test WMAP cosmology (Elvis,
\& Karovska 2002).

Near-IR interferometers are now coming on-line at the VLT (VLT-I) and
on Mauna Kea (`Ohana') with baselines of a few 100~meters. This may be
enough to resolve BEL regions, but not to image them well. A good site
for a near-IR or optical interferometer would be `Dome~C' in
Antarctica, where a new Italian-French base has been constructed. The
K-band isoplanatic patch is $\sim$1~meter in diameter at Dome~C,
allowing diffraction limited imaging with good sized telescopes quite
simply. The isoplanatic in the optical is being measured, and should
be large.  A 0.8~meter IR telescope is under construction.  In the UV,
NASA is developing an interferometry mission `Stellar Imager' with a
$\sim$20~meter baseline. The quasar community should join this effort
with a view to building a larger version, a `{\em Quasi}-Stellar
Imager', if we want to see what quasars really look like.

\begin{figure}[b]
\plotone{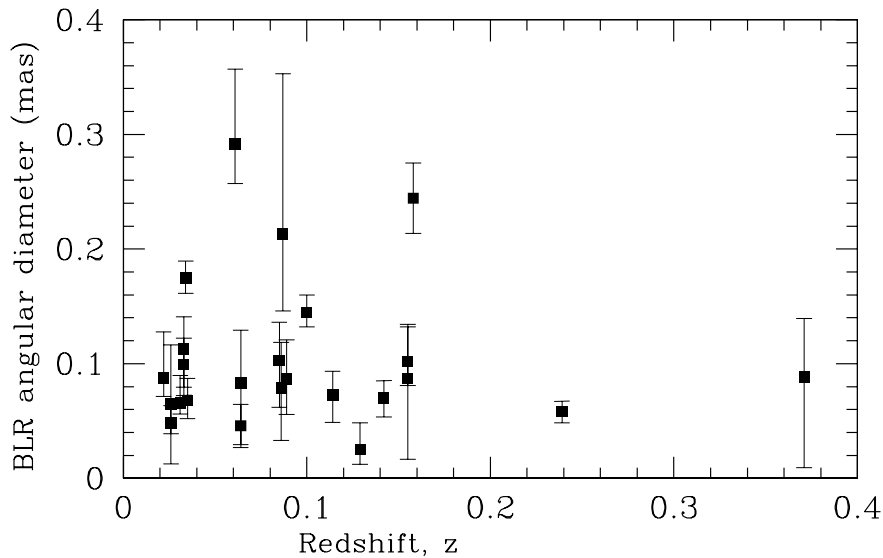}
\caption{BEL region diameters from reverberation mapping (Elvis \&
Karovska 2002).}
\end{figure}

\end{document}